\documentclass{article}
\usepackage{ijcai21}

\usepackage{times}
\usepackage{soul}
\usepackage{url}
\usepackage{bm}
\usepackage[hidelinks]{hyperref}
\usepackage[utf8]{inputenc}
\usepackage[small]{caption}
\usepackage{graphicx}
\usepackage{amsmath}
\usepackage{amsthm}
\usepackage{booktabs}
\usepackage{algorithm}
\usepackage{algorithmic}
\title{Long-term, Short-term and Sudden Event: Trading Volume Movement Prediction with Graph-based Multi-view Modeling }

\author{
Liang Zhao $^1$\thanks{\ \ Equal Contribution.} \and 
Wei Li$^2$\footnotemark[1] \and
Ruihan Bao $^3$ \and
Keiko Harimoto$^3$ \and
Yunfang Wu$^2$  \And
Xu Sun $^{1,2}$ \\
\affiliations
$^1$ Center for Data Science, Peking University \\
$^2$  MOE Key Lab of Computational Linguistics, School of EECS, Peking University  \\
$^3$  Mizuho Securities Co.,Ltd    \\
\emails
\{zhao1iang,liweitj47,wuyf,xusun\}@pku.edu.cn,
\{ruihan.bao,keiko.harimoto\}@mizuho-sc.com \\
}

\begin{document}

\maketitle

\begin{abstract}
Trading volume movement prediction is the key in a variety of financial applications.
Despite its importance, there is few research on this topic because of its requirement for comprehensive understanding of information from different sources.
For instance, the relation between multiple stocks, recent transaction data and  suddenly released events are all essential for understanding  trading market. 
However, most of the previous methods only take the fluctuation information of the past  few weeks into consideration, thus yielding poor performance.
To handle this issue, we propose a graph-based approach that can  incorporate multi-view information, i.e., long-term stock trend, short-term fluctuation and sudden events information jointly into a temporal heterogeneous graph. Besides, our method is equipped with deep canonical analysis to highlight the correlations between different perspectives of fluctuation for better prediction. Experiment results show that our method outperforms strong baselines
by a large margin.
\footnote{\ The code will be released at \url{https://github.com/lancopku/CGM}}
\end{abstract}

\section{Introduction}

Trading volume movement prediction aims to predict the volume in a certain period of time based on  stock market information, which is crucial to a variety of financial applications, e.g.,  stock market anomaly detection, risk management and algorithmic trading~\cite{volumePredictionForTradingStrategy,libman2019volumepredictionwithneuralnetworks}. 
More importantly, when investors try to buy/sell large
quantities of stocks, the order itself will instantaneously drive the stock price
to the undesirable direction (i.e., higher price for a buy order and lower
price for a sell order) and thus the total cost for the execution will be very
expensive~\cite{ForecastingtradingvolumeintheChinesestockmarketbasedonthedynamicVWAP}. Instead, if we can split the large order into smaller pieces and
execute those smaller orders according to the  market volumes movement (e.g.,
execute more when the volume is high and less when the volume is low), this
will reduce the price impact caused by the large order and thus minimize the
total execution cost for the investor. As the results, predicting the volume trend
is essential for the stock trading. 
However, volume movement prediction has been paid less attention compared with price movement prediction~\cite{DBLP:conf/acl/CohenX18,wei2020stockmovementprediction} in the AI field. 

\begin{figure*}[ht]
\centering
\small
\footnotesize
\includegraphics[width=0.99\linewidth]{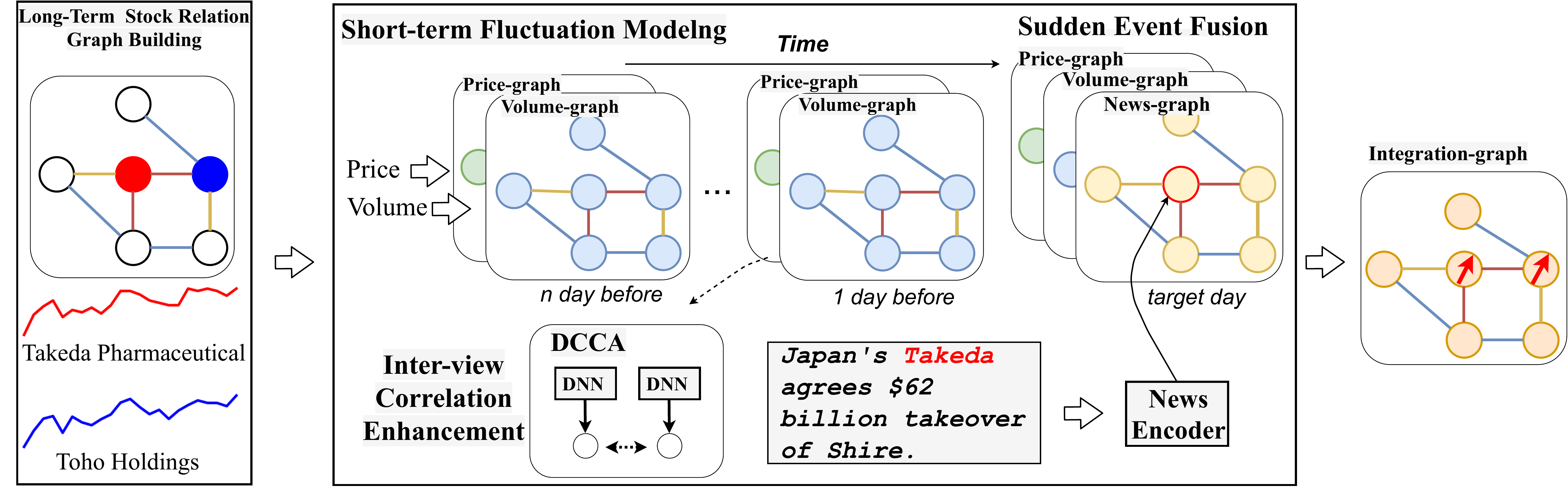}
\caption{An illustration of our proposed CGM. (1) The graph in the left most part is stock relation graph  built based on long-term price and volume data. Different colors of edges indicate different relation types. 
(Sec.\ref{sec:long_term_graph_building}). Examples of \textit{Takeda Pharmaceutical} and \textit{Toho Holdings} with long-term price trend are shown below, the color of the nodes are identical to the color of the price lines. 
(2)  
Price-graph and volume-graph are designed to process price and volume with long-term stock graph respectively.
(Sec.\ref{sec:short_term_information_processing}).  (3) Sudden events are fused with the short-term fluctuations by appending to the the end of the sequence (Sec.\ref{sec:heterogeneous_graph}). (4) Inter-view  correlation enhancement with DCCA module (Sec.\ref{sec:short_term_correlation_learning}).  The red arrow in the node of integration-graph indicates the movement trend. 
\label{fig:model}}
\end{figure*}

Previous works on volume prediction focus on predicting the volume based on short-term  transaction data. Both classic machine learning models \cite{2017Intraday,2018Point} and deep learning based models \cite{libman2019volumepredictionwithneuralnetworks} have been applied. However, even though more sophisticated models like LSTM~\cite{hochreiter1997long} have been used to model the temporal sequence, almost all the previous work considers the transaction sequence in isolation. 
Although being a valuable information resource for predicting volume movement, the randomness in the short-term transaction data largely affects the prediction accuracy. 

Therefore, introducing and integrating more comprehensive information is crucial to making more accurate predictions. We observe that there exists \textbf{long-term} connection between stocks both price-wise and volume-wise, which can be served as the regulation and facilitation for short-term information. For example, \textit{Takeda}, a Japanese pharmaceutical company, has a strong positive relation with pharmaceutical retail company \textit{Toho Holdings} on long-range price trend. 
Another observation is that \textbf{breaking news} related to a company can largely affect its trading volume. For the same stock, the news ``\textit{Japan's Takeda agree \$62 billion takeover of Shire.}'' largely increases its volume. As a result, systematical coordination of the \textbf{long-term}, \textbf{short-term} and \textbf{sudden event} information is the key for making accurate volume predictions.

Conventional methods tend to ignore long-term information~\cite{volumePredictionForTradingStrategy,2017Intraday,2018Point,libman2019volumepredictionwithneuralnetworks}, and experiments show that simply putting together information of different views cannot explicitly exploit the connection among them. 
In this paper, we propose to jointly integrate long-term, short-term and sudden event information 
into a graph via \textbf{C}orrelation-powered \textbf{G}raph-based \textbf{M}ulti-view (\textbf{CGM}) modeling method. The textual sudden information is appended to the end of the temporal sequence following short-term transaction data, which forms a temporal heterogeneous graph.
Specifically, we build a long-term relational graph through long range price and volume record information 
to model the relation among stocks with various types of edges. Then, we propose 
to encode sequential short-term transaction information together with sudden information, which combines short-term fluctuations with sudden event information by treating them together as a temporal sequence. 
We further propose to adopt deep canonical correlation analysis (\textbf{DCCA})~\cite{andrew2013deepcca} to highlight the correlation between multiple perspectives of the fluctuations.

To test the effectiveness of our approach, we collect a volume prediction dataset of Reuters Financial News, containing financial news and market data from Reuters. This dataset is to predict opening volume stock movement based on the overnight financial news~\cite{wei2020stockmovementprediction}. We conduct a series of experiments and the results demonstrate that our proposed CGM can effectively leverage multi-view information and model correlation information in transaction data points from price and volume perspectives, and thus outperforming strong baselines by a large margin.

Our main contributions are summarized as follows:
\begin{itemize}
    \item We propose and emphasize the importance of introducing long-term and sudden event information besides the previously applied short-term transaction data in the volume prediction task to regularize the randomness existing in the short period of time and detect the abrupt change of volume caused by sudden event.
    \item To systematically coordinate the multi-view information, we 
    propose a \textbf{C}orrelation-powered \textbf{G}raph-based \textbf{M}ulti-view  (\textbf{CGM}) modeling method. Heterogeneous sequential short term and textual sudden news are organized according to the multiple types of long-term stock relations in a multi-view graph. 
    
    \item Experiment results show that our method can outperform other strong baselines  by a large margin. Further analyses testify the effectiveness of the proposed multi-view modeling and  correlation learning module. 
\end{itemize}




\section{Methodology}
In this section, we first give the formulation of trading volume movement prediction. Then, we introduce the proposed model in details.

\subsection{Problem Formulation}
\label{sec:problem}
We model the trading volume movement prediction as a classification problem. Given a series of hourly transaction data points in the past few days containing $N$ data points $\bm{X}=\{x_1,x_2,\dots,x_N\}$, of which each $x_i$ consists of volume data $x^p_i$ and price data $x^v_i$. Each $x^p$ consists of hourly highest price, lowest price, open price and close price. Each $x^v$ consists of hourly volume $v$ and the proportion of it in the whole day. Besides, there are some events encoded as news information $D$ released in the market. 

Therefore, the task of volume movement prediction can be formalized as follows: given $\bm{X}$ in the past $n$ days and several financial news $D$ that take place after the trading market being closed, the model should predict the volume proportion of the first hour in the next trading day as positive or negative, which can be formally expressed as:
\begin{small}
\begin{align}
    &\bar{v} = \frac{1}{n}\sum_{i=0}^{n-1} v_i, \quad
    v_\sigma = \sqrt{\frac{1}{n-1}\sum_{i=0}^{n-1}(v_i - \bar{v})^2}, \\
    &Y = (v_{n} - \bar{v}) / v_\sigma.
\end{align}
\end{small}
To alleviate the randomness, we consider the label as positive if Y exceeds 0.5, as negative if Y is below -0.5.
We also forecast the log value of trading volume $v$ based on the same information as volume movement prediction as reference.
\subsection{Model Overview}
The overview of our method is shown in Figure~\ref{fig:model}. Our model can be partitioned into four major components, namely, a \textbf{stock relation graph} which is built by long-term stock price and volume information, 
a \textbf{short-term fluctuation module} that models price and volume data with the stock relation graph, a \textbf{sudden event fusion module} that fuses textual news information with short-term information by treating it as the last time slot of the transaction data sequence,
a \textbf{DCCA} module which is intended to 
highlight the correlation information of the price and volume 
within a single stock. 

\subsection{Long-term Stock Relation Graph Building}
\label{sec:long_term_graph_building}
The historical trading price and volume indicate long range stock attributes which can provide very valuable information for modeling relation among stocks.
To make our model consider long-term stock relation information,
we build a stock relation graph with multiple relation types based on historical trading information. In this graph, each node represents a stock. The nodes are connected according to the two correlation matrices, which are obtained by  statistical correlation (e.g., Pearson's Rank) among stocks based on the trading price and volume information in five years. 
The positive values in the graph indicate positively correlated, while negative values indicate negatively correlated. We define four kinds of relationships between nodes depending on the price correlation polarity and volume correlation polarity, namely, price (volume) positively correlated (correlation $\geq$ threshold) and negatively correlated (correlation $\leq$ threshold).

\subsection{Short-term Fluctuation Modeling}
\label{sec:short_term_information_processing}
Transaction time-series data modeling deals with the fluctuation in short term.
Because of the noisy fluctuations existed in the trading data, it is a desirable ability to incorporate long-term information to alleviate the short-term noise. 
In contrast with the previous work~\cite{libman2019volumepredictionwithneuralnetworks}, which views the price and volume data as a whole and feeds them into a neural network to get a mixed representation, we argue that there is a heterogeneity gap between price data and volume data, 
therefore we separately process price data and volume data 
as price-graph and volume-graph.

Each layer of our short-term fluctuation module consists of two parts, aggregation step and update step. The aggregation step is bound for aggregating the neighboring information, while the update step is bound for updating the hidden states.
We propose to apply RGCN~\cite{thomas2018rgcn} to model such a multi-relation graph in the aggregation step. Furthermore, to model the sequential information of the time-series data, we propose to upgrade the classic LSTM unit into a more powerful form suitable for graph as the update step in our module. At the  $l$th layer (out of $m$ layers) of the $i$th time step, the process can be expressed as follows:
\begin{small}
\begin{align}
\label{equ:gcn}
&\bm{\xi}^{l}_{i} = \sigma(\sum_{r}\bm{D}_r^{-\frac{1}{2}}\bm{A}_r\bm{D}_r^{-\frac{1}{2}}\bm{H}^{l}_{i}\bm{W}_r^{l} + \bm{W_h}\bm{H}^{l}_i) \\
&a^l_i, b^l_i, o^l_i, u^l_i, \rho^l_i = f^l_{\theta^a}, f^l_{\theta^b}, f^l_{\theta^o}, f^l_{\theta^u}, f^l_{\theta^\rho} \cdot ([\xi^{l}_{i};h^{l}_{i};x_i;h^{m}_{i-1}])\\
&c^{l+1}_i = \sigma(b^l_i)*c^l_i + \sigma(a^l_i)*\tanh(u) + \sigma(\rho^l_i)*h^m_{i-1} \\
&h^{l+1}_i = \sigma(o^l_i) \odot \tanh(c^{l+1}_i),
\end{align}
\end{small}
where $\bm{A_r}$ is the adjacency matrix of relation $r$, $\bm{D}^{-1/2}\bm{A_r}\bm{D}^{-1/2}$ denotes the normalized symmetric adjacency matrix, and $\bm{W}_{r}^{l}$ is the trainable filter in the $l$th layer for relation $r$.  
$H^l$ is the hidden representation in the $l$th layer. $\xi^{l}$ is the aggregated neighbor information in the $l$th layer.  $f_\theta$ is a one-layer feed forward network with sigmoid (denoted as $\sigma$) as activation function.  $a$, $o$, $\rho$ denote input, forget and output gates respectively. Here we use $x$ to denote price data $x_p$ or volume data $x_v$ which are taken as the input to the price-graph or volume-graph.
This module first aggregates information from neighboring nodes with aggregation layer, and then apply a gating mechanism to fuse information from the previous layer of current time-step $h^l_i$, the final layer of the previous time-step $h^m_{i-1}$, the input at current time step $x_i$ and the aggregated neighbor information $\xi^{l}_{i}$ to dynamically guide the model to select valuable information.

\subsection{Sudden Events Fusion}
\label{sec:heterogeneous_graph}
Besides short-term trading information, suddenly released events can significantly affect trading market. 
Therefore, we propose to incorporate sudden events into short-term fluctuations by appending the sudden events to the last time point of the short-term sequence, which forms a temporal heterogeneous graph.
We extract headlines of financial news to represent events, which is encoded via a news encoder.

\medskip\noindent\textbf{News Encoder}\hspace{1em}We employ a LSTM with attention mechanism~\cite{bahdanau2014neural} to extract features of news, which can be formulated as follow:
\begin{small}
\begin{align}
&\hat{h}_t = LSTM(\hat{h}_{t-1}, z_{t}),\\
&\bar{h} = Attention(e;\bm{\hat{h}};\bm{\hat{h}})
\end{align}
\end{small}
where $z_t$ is the embedding of the $t$th word, $e$ is stock node embedding as query to perform attention.

The node representation of  integration-graph is the combination of stock embedding $e$,  news text vector $\bar{h}$, price vector $h^p$  and volume vector $h^v$, which is used for the final prediction of the volume movement:
\begin{equation}
    g = \bm{W_1}(ReLU(\bm{W_2}([e;\bar{h};h^p;h^v]))),
\end{equation}
where $\bm{W_1}$ and $\bm{W_2}$ are parameter matrices and ReLU is an activation function. Finally, $g$ is taken as the  node representation in integration-graph to perform graph aggregation as Equation.~\ref{equ:gcn} to obtain the final  stock representation $\hat{g}$.

\subsection{Inter-view Correlation Enhancement}
\label{sec:short_term_correlation_learning}
To highlight the correlation between volume and price of a single stock from a global angle,  inspired by recent advances of several work~\cite{sun2020deepcca_multimodal,gao2020deepcca_multimodal}, we propose to use a deep canonical correlation analysis (\textbf{DCCA})~\cite{andrew2013deepcca} to capture essential common information from both price and volume perspectives. Specifically, the output of price-graph and volume-graph is passed into a Siamese
Network~\cite{koch2015siamese}, whose two bodies are denoted as $\phi$ and $\psi$, to transform the input into common space with non-linear transformations. DCCA attempts to maximize the correlations of the two perspectives via maximizing the output of $\phi$ and $\psi$, denoted as $F_X = \phi(X^p;\theta_1)$ and $F_Y = \psi(X^v; \theta_2)$, by finding two linear transformations $C^*$, $D^*$. The objective of DCCA is expressed as follows:
\begin{equation}
\small
\begin{split}
    \theta^*_\phi, \theta^*_\psi &= \mathop{\arg\min}_{\theta_\phi, \theta_\psi}  \rm{CCA}(F_X, F_Y)\\
    &=\mathop{\arg\max}_{\theta_\phi, \theta_\psi} \rm{corr}(C^{*T}F_X,D^{*T} F_Y).
\end{split}
\end{equation}
Let $R_{11}$, $R_{22}$ be covariances of $F_X$, $F_Y$, the cross-covariance of $F_X$, $F_Y$ as $R_{12}$. Let $E = R^{\frac{1}{2}}_{11}R_{12}R^{\frac{1}{2}}_{22}$, and the canonical correlation loss for updating $F_x$, $F_y$ can be defined as
\begin{equation}
\label{equ:cca_loss}
   loss^{corr} = -\rm{trace}(E^TE)^{\frac{1}{2}}.
\end{equation}
We minimize $loss^{corr}$ to highlight correlation between price and volume data of a single stock.

\subsection{Objective}
Given the aforementioned structure, the entire volume movement prediction process can be formalized as follows:
\begin{small}
\begin{align}
    &\bm{P} = softmax(\bm{W_l}\hat{\bm{g}}) \\
    \label{equ:class_loss}
    &loss^c = -\sum q log(\bm{P}). \\
    \label{equ:loss}
    &loss = loss^{c} + \lambda loss^{corr}. 
\end{align}
\end{small}
Where $q$ is the volume movement label. $\lambda$ is a coefficient to balance two parts of loss. $W_l$ is a parameter matrix.
\section{Experiment}
This section describes the experimental dataset, evaluation metrics and baseline algorithms for comparison.
\subsection{Dataset}
We choose the top 500 stocks in Tokyo Exchange known as TPX500 to perform experiments.
Because the news data may contain noise which do not affect the stock market, following~\cite{wei2020stockmovementprediction}, we first pick out the news with the ``RIC'' label provided in the data by Reuters, indicating it to be a stock-influencing news. Then we pick out the news with financial keywords provided by~\cite{chen2019stock_movement_prediction}. 
In this task, we only predict the movement when both the news is available and the volume movement surpasses 0.5 times of hourly standard deviation after subtracting the mean of hourly volume of this stock. Finally, there are 27,159 positive movements and 34,562 negative movements. We split the data in the period of 01-01-2018 $\sim$ 04-30-2018 as the development set and the data in the period of 05-01-2018$\sim$09-30-2018 as the test set and the rest as training data. More detailed statistical information of the dataset is demonstrated in Table~\ref{tab:dataset}.

\subsection{Evaluation Metrics}
For volume movement prediction, we use \textbf{\textit{accuracy}} to evaluate the model performance. Besides, we also perform log volume value prediction to further verify the performance of each model as reference. We minimize a MSE loss in place of the cross-entropy loss in Equ.~\ref{equ:class_loss} for regression. In this task, we adopt  \textbf{MSE} = $\frac{1}{M}\sum_{i=1}^{M} \sqrt{(y_i - \hat{y}_{i})^2}$ and \textbf{RMSE} = $\sqrt{\frac{1}{M}\sum_{i=1}^{M} (y_i - \hat{y}_{i})^2}$  to evaluate the errors between ground-truth log volume and predictive output.

\subsection{Baseline Models}
We include several algorithms as baselines to compare the performance, including classic methods as well as state-of-the-art models based on deep neural networks.
\begin{itemize}
\item \textbf{Random}: this model predicts the movement to improve or decline randomly.
\item \textbf{Moving Average (MA)}: takes the arithmetic mean of the past $n$ day volume data as prediction.
\item Classic classification and regression methods: \textbf{Random Forest (RF)}, \textbf{Logistic Regression (LogisticR)}, \textbf{Linear Regression (LinearR)}, \textbf{Support Vector Machine (SVM)}, \textbf{LSTM}.
\item \textbf{LSTM+SVM}~\cite{libman2019volumepredictionwithneuralnetworks}: is a hybrid model combining the results from LSTM and SVM, which uses the LSTM output as the SVM feature to do prediction. 
\item \textbf{LSTM-RGCN}~\cite{wei2020stockmovementprediction} is a graph-based method with LSTM mechanism to alleviate over-smoothing. In their method, news and stock relation information are taken into consideration.
\item \textbf{FinBERT}~\cite{Araci2019finbert} is a language model pre-trained on financial news. Here we finetune the released model with the news in our dataset for prediction.
\end{itemize} 

\subsection{Settings}
 For machine learning baselines (RF, LogisticR, LinearR, SVM), we use the implementation and default setting in scikit-learn\footnote{\url{https://scikit-learn.org/stable/}}. For other baselines, we use the released code and default settings for implementation. The term $n$ is set as 20, the $\lambda$ in loss function is empirically set as 1. The siamese network is implemented with a three-layer MLP~\cite{benjio2009mlp} and the two bodies do not share parameters. For LSTM and LSTM+SVM, we use the transaction data of price and volume in the past 20 days as input feature. For other baselines, we concatenate price and volume  hourly data in the past 20 day as input to do prediction. The layer number of LSTM and LSTM+SVM is set to be 2. Our price-graph and volume-graph have one aggregation layer and one update layer and integration-graph has one aggregation layer. We set the threshold
of correlation edge to 0.6, that is, only when the weight
of the edge exceeds 0.6, there is an edge built between the
two nodes. We use GLoVe~\cite{glove2014emnlp} trained on financial news as the initialization of word embedding, and the size of word embedding is 50. The hidden size of LSTM-based baseline and our model is 300. We use Adam optimizer to train LSTM-based baselines and our model. The learning rate is in the range of $\{10^{-6}, 10^{-5}, 10^{-4}, 10^{-3}, 3\times10^{-3}\}$ and we adopt models with the best performance in developing set for testing for each baseline and the proposed CGM.
\begin{table}[t!]
  \centering
  \small
  \scalebox{1}{\begin{tabular}{lcccc}
  \toprule  
  \textbf{Dataset} & \textbf{Node \#} & \textbf{Train} & \textbf{Dev} & \textbf{Test}   \\
  \midrule  
   TPX500  & 498 & 53,994 & 3,310 & 4,417 \\
  \bottomrule 
  \end{tabular}}
  \caption{Dataset statistics. ``Node \#'' is the number of the stocks of the dataset. ``Train'', ``Dev'', and ``Test'' are the sizes of the training set, development set, and test set.  }
  \label{tab:dataset}
  \end{table}


\begin{table}[t]
\centering
\small
\begin{tabular}{|c|c|c|c|}
\hline Methods & Accuracy$\uparrow$ & $\mathrm{MSE}$$\downarrow$ & $\mathrm{RMSE}$$\downarrow$ \\
\hline           Random & 50.12 $\pm$ 0.76 &  * & *  \\
  LogisticR & 56.26 $\pm$ 0.00 &  * & *   \\
  LinearR & * & 0.1699 & 0.4122  \\ 
  MA & * & 0.2033 & 0.4509 \\
  RF & 54.97 $\pm$ 1.68 &  0.1835 & 0.4283  \\
  SVM & 53.91 $\pm$ 0.00 & 0.1755 & 0.4189   \\

  LSTM & 57.65 $\pm$ 2.44 & 0.1706 & 0.4131   \\
  LSTM+SVM & 57.13 $\pm$ 2.13 & 0.1657 & 0.4070  \\
  LSTM-RGCN & 53.38  $\pm$ 2.57 & * & * \\
  FinBERT & 54.50 $\pm$ 3.26 & * & * \\
\hline \hline 
CGM (ours) & \textbf{61.21} $\pm$ 2.32 & \textbf{0.1611}  & \textbf{0.4013}    \\
-Improvement & \textbf{+6.2\%} & \textbf{+2.7\%} & \textbf{+1.4\%}  \\
\hline
\end{tabular}
\caption{Overall performance of all the models.  \textit{Accuracy} is the higher the better, MSE and RMSE scores are the lower the better. The best performance is highlighted in \textbf{bold}. We show average values of five experiments with different random seeds for each model in all the three metrics. In volume movement prediction task,  we also show the standard deviation of all the models. }

\label{table:experiment_results}
\end{table} 

\begin{table}[t]
\centering
\resizebox{.9\columnwidth}{!}{
\begin{tabular}{@{}l|c|c|c@{}}
\toprule \textbf{Model Variants} & \textbf{Acc}$\uparrow$ & $\textbf{MSE}$$\downarrow$ & $\textbf{RMSE}$$\downarrow$ \\
\midrule
  Full Model & \textbf{61.21}  &  \textbf{0.1611} & \textbf{0.4013}  \\
\hline 
$\bm{w/o}$ news & 58.03 ($\downarrow$3.18)   &  0.1624 &  0.4029  \\
$\bm{w/o}$ DCCA & 59.18 ($\downarrow$2.13) & 0.1671 & 0.4088 \\
$\bm{w/o}$ integration-graph  & 58.68 ($\downarrow$2.53) & 0.1680 & 0.4092 \\
$\bm{w/o}$ volume-graph & 56.26 ($\downarrow$4.95) &  0.1693 & 0.4115 \\
$\bm{w/o}$ price-graph & 59.36 ($\downarrow$1.85) & 0.1684 & 0.4097\\

\bottomrule  
\end{tabular}}
\caption{Ablation studies. ``$\bm{w/o}$ news'' denotes that we don't introduce news  into integration-graph. ``$\bm{w/o}$ DCCA'' denotes that we assign $\lambda$ in Eqn~\ref{equ:loss} to zero. ``$\bm{w/o}$ integration-graph '' means that we combine news and output of price graph and volume graph, and then feed them to a two-layer MLP to do prediction. ``$\bm{w/o}$ volume/price-graph'' denotes replacing volume/price graph with a LSTM neural network to model short-term time-series data. }
\label{table:ablation_study}
\end{table} 

          

\subsection{Results}
The overall performance is shown in Table~\ref{table:experiment_results}.
There are several major observations. First, random guess yields the weakest result which is about 50\%, and our proposed CGM outperforms all the baselines by a large margin, exceeding the best LSTM-based baseline by \textbf{6.2\%} improvement, which confirms the validity of our proposed model.
Second, we can find that simple baseline LR shows competitive results  compared with deep learning based baselines but far from our method, we argue that the reason  may be that the increasing of  model complexity is not the key of performance improvement, instead, the introduction of different views (long-term and sudden information) is crucial to  making comprehensive prediction, therefore yielding better performance.
Third, on regression metrics MSE and RMSE, our proposed method also exceeds other baseline models, showing that our model is not only capable of identifying the movement of trading volume but also can forecast specific value of it more precisely. The reason behind
may be that our approach can better identify the direction of the stock market shock by capturing more comprehensive information, which makes it able to make a more accurate regression forecast. Fourth, compared with \textit{accuracy},  the gaps on MSE scores and RMSE scores of different methods are smaller. We assume that trading volume regression is a more challenging task which requires models to forecast the specific value of volume, however there is a lot of randomness in the specific value of volume and predicting the movement of volume is a more feasible objective.

\subsection{Analyses}

\subsubsection{Effect of Multi-view Integration}
To test the effect of integration-graph, we replace the integration-graph with a two-layer MLP to combine sequential short-term information and sudden event information. More specially, short-term transaction data is encoded with a linear LSTM, the output of which is concatenated with the output of news encoder followed by a MLP network.
From the result in Table \ref{table:ablation_study} we can see that the accuracy significantly declines by $2.53$ points. Besides, regression based metrics also suffer big loss.
The reason we assume is that the long-term relation information within the graph is beneficial for comprehensively integrating different views of market information and regularizing the randomness existed in short-term fluctuations. Besides, the graph structure can make useful information propagate from a node to its neighbours, therefore contributing to forecasting for these nodes.

\subsubsection{Effect of Sudden Event}
To test the effect of introducing sudden event information, we remove the news encoder in our module.
We can find that when news information is removed, the performance degradation of our method is obvious on \textit{accuracy} but not as remarkable on MSE and RMSE scores. This shows that news information can provide big help to model sudden accident which is important to detect stock market movement. However, the effect is limited for volume value forecasting, indicating that the volume value forecasting relies more on recent transaction variation, and it is a challenging problem to better exploit textual information for digital related regression, which needs further research. 

In the right part of Figure \ref{fig:news_mask}, we show the results of different filtering standard deviations defined in section \ref{sec:problem}. It can be seen that news information brings bigger influence under more significant movement situations, which is reflected by the fact that models using news (proposed CGM model and FinBERT) gain more improvement than the baseline model that only involves transaction data. Under both circumstances, our model enjoys consistent better performance than baseline models. Furthermore, we can observe that when the movement is larger, the traditional method will suffer more performance degradation. This phenomenon testifies that using only short-term information is insufficient for volume movement forecasting.

\subsubsection{Effect of Long-term Graph}
To see the effect of long-term relation graphs, we separately remove price and volume graph. From the results we can see that both kinds of graph have positive effect on volume prediction. It can be also observed that volume-graph has larger effect than price-graph, this is expected because long-term volume information has more direct connection with volume movement. On the other hand, price-graph also contributes to the prediction of volume, which shows the importance of multi-view modeling.

\subsubsection{Effect of DCCA}
When the DCCA module is removed, both regression and classification scores decline, verifying the effectiveness of highlighting correlations between different perspectives of the volume and price of a single stock. This phenomenon testifies our assumption that the correlated part between price and volume information reveals the movement trend of the stock while alleviating the effect of noise in the price or volume individually. Furthermore, this shows that the simple concatenation of price and volume cannot effectively model the correlation between price and volume.

\subsubsection{Associative Prediction Analysis}
\begin{figure}[t]
\centering
\includegraphics[width=0.99\linewidth]{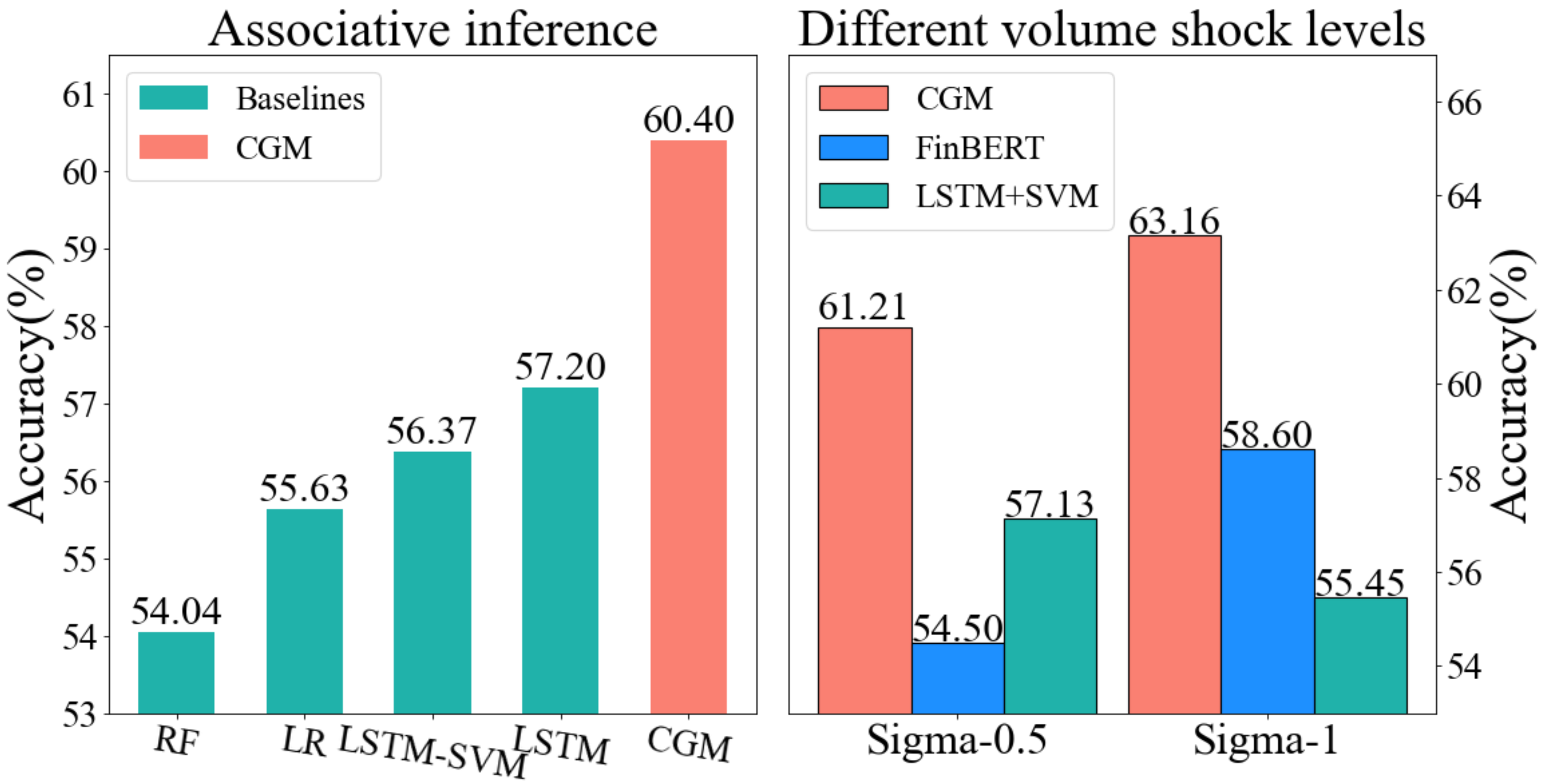}
\caption{(a) The left part of the figure is the associative stock movement inference result, where the nodes are not attached with news. (b) The right part is the results of using different standard deviations to filter movements regarding to the test dataset construction in section \ref{sec:problem}. It can be seen that our model performs consistently better than baselines with different volume shock levels. 
\label{fig:news_mask}}
\end{figure}

There are many situations where stocks are not associated with news which we denote as no-news nodes.  Because of the graph structure, our method can cope with this situation by the propagation of news information from news-associated stocks to their no-news neighboring stocks via graph edges. Figure \ref{fig:news_mask} shows the association inference results.
We can find that our method  surpasses other baseline models by a large margin. This verifies our conjecture that our proposed graph structure is effective and the news information can efficiently propagate from news-associated stocks to \textit{no-news} stocks, which is crucial to forecasting the shock in trading volume for those stocks. 

\section{Related Work}
\subsection{Volume Prediction}
Traditional volume prediction focuses on predicting the volume with classic machine learning methods solely based on preceding transaction data. Liu \textit{et al.}~\shortcite{2017Intraday} propose to apply support vector machine for regression to predict the changes of volume percentage. Roman~\shortcite{2018Point} propose to apply Bayesian inference to model and forecast intraday trading volume, using auto-regressive conditional volume (ACV) models.
With the rapid development of deep learning, researchers have also converted to apply neural network models to predict volume. Libman \textit{et al.}~\shortcite{libman2019volumepredictionwithneuralnetworks} propose to apply LSTM to model the time series transaction data followed by the support vector machine for regression to predict the trading volume. Oliveira \textit{et al.}~\shortcite{oliveira2017impact} attempt to use textual information from micro-blog, but the text in their work only plays the role of the source of sentiment. It is the sentiment result rather than the text that is used to do prediction. In this paper, we introduce multi-view information 
to better understand stock market for volume prediction.

\subsection{Heterogeneous Information Integration}
Comprehensively understanding of information from multiple sources is paramount to improving the performance of volume prediction. However, the heterogeneity gap between information sources makes it hard to integrate heterogeneous information. Previous researches mainly focus on mapping multimodal data into the common space. \cite{yang2019heterogeousfusion} propose to utilize GCN~\cite{gcn2017} to combine multi-aspect information of entities to learn the entity embeddings of multilingual knowledge graphs. \cite{DBLP:conf/emnlp/ZadehCPCM17} propose to integrate language, audio and video data into a common space via tensor outer product operation. Other researchers attempt to measure the distance between two modalities, which is minimized during training \cite{DBLP:journals/tmm/LiongLTZ17a}. Generative models like auto-encoder and GAN have also been applied to learn joint cross modality representations \cite{DBLP:conf/ijcai/WangCO015,DBLP:journals/tomccap/PengQ19}. More recent works introduce  models pre-trained on large cross-modality corpus. \cite{DBLP:conf/eccv/ChenLYK0G0020} propose several modality specific and cross modality pretraining objectives to train a unified Transformer model for join image-language representation. Although these previous work provides good insight for us to integrate heterogeneous information from different sources to predict volume movement, our task input is quite different from traditional tasks where the input is usually image-language or video-language. The connection between news text and transaction data is more flexible and trivial. Therefore, we propose a graph based module where the graph is constructed out of multiple sources and we further propose a deep canonical correlation analysis module to enhance the connection between different aspects of the short-term transaction data of a single stock.

\section{Conclusion}
In this paper, we propose a novel temporal heterogeneous graph that can integrate multi-view information, including long-term stock relation, short-term fluctuations and sudden events for trading volume movement forecasting.
In addition, our proposed method is equipped with a DCCA module to highlight the correlation of transaction information between volume and price from a global angle.
Experiment results  verify the effectiveness of our proposed method and the integration of information from different views. 
Our method outperforms other strong baselines by a large margin on this task. Further analyses attest that  the introduction of news indeed contributes to the prediction accuracy, and our model performs consistently better than baselines at different volume shock levels.


\section*{Acknowledgments}
The authors would like to thank 
responsible reviewers of IJCAI21 for their 
thoughtful and
constructive suggestions. This work is partly supported  by Beijing Academy of Artificial Intelligence (BAAI). Xu Sun is the corresponding author.
\appendix
\bibliographystyle{named}
\bibliography{ijcai21}

\end{document}